\documentclass[a4paper,conference]{IEEEtran}
\ifCLASSINFOpdf
\else
\fi
\usepackage[cmex10]{amsmath}
\usepackage{amsmath}
\usepackage{tikz}
\usetikzlibrary{arrows,snakes,shapes,positioning,arrows,decorations.markings,calc}
\usetikzlibrary{dsp,chains}
\usepackage{epsf}
\usepackage{epsfig}
\usepackage{psfrag}
\usepackage{graphicx}

\IEEEoverridecommandlockouts
\IEEEpubid{\makebox[\columnwidth]{978-1-4673-6540-6/15/\$31.00~\copyright~2015 IEEE \hfill} \hspace{\columnsep}\makebox[\columnwidth]{ }}

\begin{document}

\title{Power- and Spectral Efficient Communication System Design Using 1-Bit Quantization}

\author{\IEEEauthorblockN{Hela Jedda, Muhammad Mudussir Ayub, Jawad Munir, Amine Mezghani and Josef A. Nossek}
\IEEEauthorblockA{Institute for Circuit Theory and Signal Processing\\
Munich University of Technology, 80290 Munich, Germany\\
E-Mail: \{hela.jedda, josef.a.nossek\}@tum.de}
}

\maketitle

\tikzset{DSP lines/.style={help lines,very thick,color=black}}
\tikzset{line_arrow/.style={help lines,very thick,color=black,->,-angle 90}}
\tikzset{filter/.style={rectangle,inner sep=0pt,minimum height=0.8cm,minimum width=1cm,draw=black,very thick}}
\tikzset{delay/.style={rectangle,inner sep=0pt,minimum size=1cm,draw=black,very thick}}
\tikzset{downsampling/.style={rectangle,inner sep=0pt,minimum height=0.8cm,minimum width=0.7cm,draw=black,very thick}}
\tikzset{upsampling/.style={rectangle,inner sep=0pt,minimum height=0.8cm,minimum width=0.7cm,draw=black,very thick}}
\tikzset{empty_node/.style={inner sep=0pt,minimum size=0cm}}
\tikzset{connection/.style={circle,draw=black,fill=black,inner sep=0pt,minimum size=2mm}}
\tikzset{coefficient/.style={isosceles triangle,draw=black,very thick,inner sep=0pt,minimum size=.7cm}}
\tikzset{source/.style={semicircle,minimum size=.5cm,draw=black,very thick,shape border rotate=270}}
\tikzset{adder/.style={circle,minimum size=.25cm,inner sep=0pt,draw=black,very thick}}
\tikzset{multiplier/.style={circle,minimum size=.25cm,inner sep=0pt,draw=black,very thick}}
\tikzset{double_arrow/.style={double distance=5pt,thick,shorten >= 6pt,decoration={markings,mark=at position 1 with {\arrow[scale=.6,>=angle 90]{>}}},postaction={decorate}}}



%
\IEEEpeerreviewmaketitle

\begin{abstract}
Improving the power efficiency and spectral efficiency of communication systems has been one of the major research goals over the recent years. However, there is a trade-off in achieving both goals at the same time. In this work, we consider the joint optimization of the power amplifier and a pulse shaping filter over a single-input single-output (SISO) additive white Gaussian noise (AWGN) channel using $1$-bit analog-to-digital (ADC) and digital-to-analog (DAC) converters. The goal of the optimization is the selection of the optimal system parameters in order to maximize the desired figure-of-merit (FOM) which is the product of power efficiency and spectral efficiency. Simulation results give an insight in choosing the optimal parameters of the pulse shaping filter and power amplifier to maximize the desired FOM.
\end{abstract}

\begin{IEEEkeywords}
Green Communication, Power Efficiency, Spectral Efficiency, 1-Bit Quantization.
\end{IEEEkeywords}

\section{Introduction}
\label{sec:intro}
The demand for higher throughput and capacity is increasing at a staggering rate per year for cellular networks. 
Higher cell density, greater amount of usable spectrum and higher spectral efficiency are considered as key possibilites 
to realize such an increase in capacity \cite{Rangan2014}. The deployment of very large number of antennas, also known as massive MIMO, at 
the base station is considered as key enabler to meet very high data requirements.

One of the major limiting factors for the implementation of massive MIMO are the complexity issues and the power 
consumption of the RF components due to the large number of antennas. Several approaches are considered in the literature 
to decrease the power consumption such as spatial modulation \cite{Renzo2014}, the use of parasitic antennas \cite{Kalis2012} and the use of low-cost transceivers \cite{Bjornson2015}. One attractive 
solution is the use of very low resolution ADCs and DACs, because the power consumption increases exponentially with resolution \cite{Svensson2006}. 
The extreme case of $1$-bit ADCs and DACs is particularly very interesting because it can greatly simplify and decrease the 
power requirements of the other RF front-end components such as power amplifiers (PA), low-noise amplifiers (LNA) etc. 

The price to pay for the $1$-bit is the severe signal distortion due to the strong non-linear operation. It also causes both 
in-band and out-of-band radiation (OBR) resulting in the degradation of the bandwidth efficiency. Therefore, the potential advantages of $1$-bit 
systems have to be supported by new signal processing algorithms and RF front-end architectures to achieve bandwidth efficiency and spectral 
efficiency. In this paper, we investigate the effect of pulse shaping and optimum operating region of the power amplifier for a $1$-bit DAC and ADC 
at the transmitter and receiver side, respectively. As a basic investigation of possible receiver and transmitter structures, we consider the SISO case and we leave the MIMO case for future investigation. 

The paper is organized as follows: First, the SISO system model is introduced in Section \ref{sec:system_model}. Section \ref{sec:power_model} describes the general power model of different components in a wireless transceiver front-end. Performance metrics and the optimization problem are given in Section \ref{sec:performance_measures} and Section \ref{sec:problem_formulation}, respectively. Section \ref{sec:simulation_results} analyzes the optimal system parameters that solve the optimization problem.

\section{System Model}
\label{sec:system_model}
%
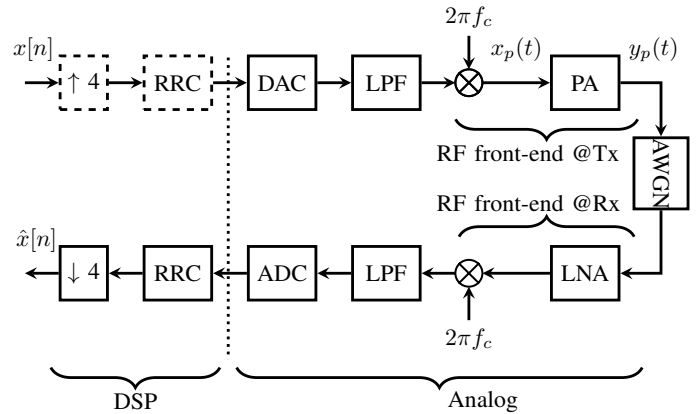
\begin{figure}[h]
\centering
\resizebox{9cm}{!} {%
\begin{tikzpicture}

\node (in){};
\node [right=of in] [xshift=-1.3cm,yshift=0.5cm]{$x[n]$};
\node[upsampling] (upsampling) [right=of in][xshift=-0.5cm, dashed] {$\uparrow$  4};
\node[filter] (pulseshaper) [right=of upsampling] [xshift=-0.5cm, dashed]  {RRC};
\node (n1)[right=of pulseshaper] [xshift=-0.9cm, yshift=0.5cm] {};
\node[filter] (dac) [right=of pulseshaper] [xshift=-0.5cm] {DAC};
\node[filter] (optical) [right=of dac] [xshift=-0.5cm]  {LPF};
\node (br1_r) [below=of optical][xshift=0.9cm,yshift=1cm] {};
\node[right=of br1_r][xshift=-1.4cm,yshift=-0.5cm]{RF front-end @Tx};
\node [dspmixer] (mod) [right=of optical][xshift=-0.5cm]   {};
\node (mod_f) [above=of mod] [yshift=-0.5cm]{$2\pi f_c$};
\node [right=of mod][xshift=-1cm,yshift=0.5cm] {$x_p(t)$};
\node [filter] (pa) [right=of mod]   {PA};
\node [right=of pa][xshift=-1cm,yshift=0.5cm] {$y_p(t)$};

\node (br1_l) [below=of pa][xshift=0.9cm,yshift=1cm]{};

\node[filter]  (adder2)  [right=of pa][xshift=-0.4cm,yshift=-0.8cm, rotate=-90] {AWGN};
\node [filter] (lna) [below=of pa][yshift=-1cm]{LNA};
\node (br2_r) [above=of lna][xshift=0.6cm,yshift=-1cm] {};
\node[dspmixer] (demod) [left=of lna] {};
\node (demod_f) [below=of demod][yshift=0.5cm] {$2\pi f_c$};
\node[filter] (receiver) [left=of demod][xshift=0.5cm]    {LPF};
\node (br2_l) [above=of receiver][xshift=1.2cm,yshift=-1cm]{};
\node[right=of br2_l][xshift=-1.7cm,yshift=0.5cm]{RF front-end @Rx};
\node [dspmixer] (mod) [right=of optical][xshift=-0.5cm]   {};
\node [filter] (adc) [left=of receiver]  [xshift=0.5cm] {ADC};
\node (n2)[below=of n1] [yshift=-3.5cm] {};
\node (n3) [left=of n2][xshift=-1.5cm] {};
\node (text_dsp) [left=of n2][xshift=0.2cm, yshift=-0.5cm] {DSP};
\node (n4) [right=of n2][xshift=5cm] {};
\node (text_dsp) [right=of n2][xshift=2cm, yshift=-0.5cm] {Analog};
\node [filter] (eq) [left=of adc][xshift=0.5cm]   {RRC};
\node[downsampling] (downsampling) [left=of eq][xshift=0.5cm] {$\downarrow$ 4};
\node (out) [left=of downsampling][xshift=0.5cm]  {};
\node [left=of downsampling] [xshift=1.1cm,yshift=0.5cm]{$\hat x[n]$};

\draw[DSP lines] [-stealth] (in.east) -- (upsampling.west);
\draw[DSP lines] [-stealth] (upsampling.east) -- (pulseshaper.west);

\draw[DSP lines] [-stealth] (pulseshaper.east) -- (dac.west);
\draw[DSP lines] [-stealth] (dac.east) -- (optical.west);
\draw[DSP lines] [-stealth] (optical.east) -- (mod.west);
\draw[DSP lines] [-stealth] (mod.east) -- (pa.west);
\draw[DSP lines] [-stealth] (pa.east) -| (adder2.west);
\draw[DSP lines] [-stealth] (adder2.east) |- (lna.east);
\draw[DSP lines] [-stealth] (lna.west) -- (demod.east);
\draw[DSP lines] [-stealth] (demod.west) -- (receiver.east);
\draw[DSP lines] [-stealth] (receiver.west) -- (adc.east);
\draw[DSP lines] [-stealth] (adc.west) -- (eq.east);
\draw[DSP lines] [-stealth] (eq.west) -- (downsampling.east);
\draw[DSP lines] [-stealth] (downsampling.west) -- (out);

\draw[DSP lines] [-stealth] (mod_f) -- (mod.north);
\draw[DSP lines] [-stealth] (demod_f) -- (demod.south);

\draw [very thick, dotted] (n1) -- (n2);
\draw [very thick, decoration={brace, amplitude=0.3cm}, decorate] (n2)--(n3);
\draw [very thick, decoration={brace, amplitude=0.3cm}, decorate] (n4)--(n2);
\draw [very thick, decoration={brace, amplitude=0.3cm}, decorate] (br1_l.west) -- (br1_r.east);
\draw [very thick, decoration={brace, amplitude=0.3cm}, decorate] (br2_l.west) -- (br2_r.east);

\end{tikzpicture}%
}
\caption{Wireless Communication System Design: sys1 (infinite resolution in DAC/ADC), sys2 (1-bit DAC/ADC) and sys3 (1-bit DAC/ADC w/o RRC and upsampling @ Tx)}
\label{fig:sys_model}
\end{figure}
	
Consider the block diagram of a communication system transmitting QPSK symbols generated at a baud rate of $B$ depicted in Fig. \ref{fig:sys_model}. In the transmitter, the QPSK symbols are shaped by a discrete-time root-raised cosine (RRC) filter. The resulting baseband signal is converted to the analog domain, utilizing two DACs for the inphase and quadrature components. The low-pass filter (LPF) removes the DACs' alias spectra. The baseband signal is then translated to the RF carrier frequency by the upconversion mixer. The PA generates the required output power of the signal and delivers it to the antenna. The transmitted signal of power $P_{\text{T}}$ propagates through an AWGN channel between the transmitter and receiver in the form of an electromagnetic wave and gets disturbed by some Gaussian distributed noise of power spectral density $N_0$. The signal-to-noise ratio (SNR) is then defined as $\text{SNR} = \alpha \frac{P_{\text{T}}}{\sigma^2_n}$, where $\alpha$ represents the constant channel power gain and $\sigma^2_n = N_0 B$. The signal captured by the receiver antenna is amplified by an LNA and converted into a baseband signal and then filtered by a LPF to limit the noise bandwidth. In this work, an ideal LNA will be considered since its power consumption is in general quite negligible \cite{mezghani2011modeling}. The baseband signal is converted into discrete-time using two ADCs for the inphase and quadrature components. After the ADCs, the RRC filter of the receiver completes the pulse shape to fulfill the Nyquist inter-symbol-interference (ISI) criterion and the sampling rate is reduced to the baud rate $B$. In this work, three system models are considered. Sys1 designates the above described system with infinite resolution DAC/ADC. Sys2 is the same system model but with restriction of 1-bit DAC/ADC. Sys3 differs from sys2 through the removal of RRC and the upsampling at the transmitter side. In this case, the pulse shaping is realized by the LPF and the band pass filter (BPF) of the PA.
 
The total power consumed by the communication system $P_{\mathrm{Total}}$ can be divided into two major parts: power consumed by the analog front-end components $P_{\mathrm{Analog}}$ and power of digital signal processing $P_{\mathrm{DSP}}$:
\begin{equation}\label{eq_1}
  P_{\mathrm{Total}}= P_{\mathrm{Analog}} + P_{\mathrm{DSP}}.
\end{equation}
The aim of this contribution consists in reducing $P_{\mathrm{Analog}}$. The power consumed by the analog front-end components $P_{\mathrm{Analog}}$ can be given as:
\begin{equation}\label{eq_2}
  P_{\mathrm{Analog}}=  P_{\mathrm{PA}} + P_{\mathrm{DAC}} + P_{\mathrm {ADC}},
\end{equation}
where $P_{\mathrm{PA}}$, $P_{\mathrm{DAC}}$ and $P_{\mathrm{ADC}}$ are the powers consumed by the PA, DAC and ADC, respectively.

To this end, these three components will be investigated in more detail.

\section{Power Model}
\label{sec:power_model}
\subsection{Power Amplifier (PA)}
The PA is a non-linear device which clips the output signal if the input signal is greater than a certain value. Non-linearities of the PA cause interferences and distortions, which create in-band and out-of-band spectral spread \cite{morgan2006generalized}. To quantify $P_{\mathrm{PA}}$, a model of PA has to be first introduced and then a closed form for $P_{\mathrm{PA}}$ calculation can be given.
\subsubsection{Modeling and Implementation}
\begin{figure}[!]
                \centering
                \psfrag{RL}[][]{$\!\!R_{L}$}
                \psfrag{iL}[][]{$i_{L}(t)$}
                \psfrag{ut}[][]{$v_{t}(t)\!\!\!\!\!\!\!\!$}
                \psfrag{id}[][]{$\!\!\!\!i_{d}(t)$}
                \psfrag{U0}[][]{$\!\!\!\!V_{sat}$}
                \psfrag{U1}[][]{$V_{th}\!\!\!\!$}
                \psfrag{x}[][]{$\!\!\!\!\!\!\!\!x_{p}(t)$}
                \psfrag{yt}[][]{$y_{p}(t)\!\!\!\!\!\!\!\!$}
                \psfrag{bpf}[][]{$h_{bpf}(t)$}
                \epsfig{file=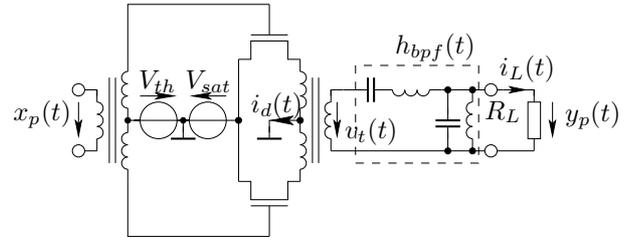, width = 0.8\columnwidth}
                \caption{Simplified push-pull transformer amplifier}
                \label{fig:push_pull}
      \end{figure}

\begin{figure}
\centering
\psfrag{A}[][]{$A/V_{sat}$}
\psfrag{f(A)}[][]{$f(A)/V_{sat}$}
\epsfig{file= 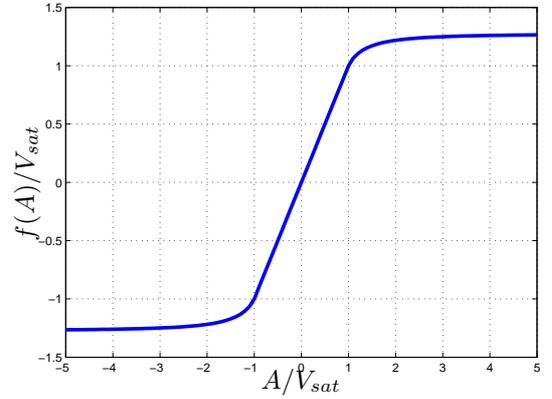, width = 0.8\columnwidth}
\caption{PA AM-AM transfer characteristic: $f(A)/V_{sat}$ vs. $A/V_{sat}$}
\label{fig:PA_AM_AM}
\end{figure}

The PA is a critical component in wireless communication systems and correct modeling of its non-linearities are of utmost importance. The dominant PA distortion source is the amplitude distortion or AM-AM conversion. It describes the relation between the amplitudes of the PA's input and output signals \cite{ali2008behavioral}. The AM-AM distortion can incorporate most of the PA non-linear effects \cite{liu2005practical}. The PA is modeled as a transformer based push-pull stage as depicted in Fig. \ref{fig:push_pull}, where $V_{sat}$ is the supply voltage, $V_{th}$  is the bias voltage for the transistor, $i_{L}(t)$ is the load current and $R_{L}$ is the load resistance of the antenna. This model is called push-pull because only one transistor is on at a time. It behaves as a voltage controlled current source if the input voltage is small and for higher input values it starts behaving as a switch. For a given PA input signal $x_{p}(t) = A(t)\cos(2 \pi f_{c}t + \phi(t))$, under single carrier and narrow band assumption, the output signal $v_t(t)$, which is affected by the AM-AM distortion, can be given by:
\begin{align}
v_{t}(t)&= \mathrm{sign}\left(\cos\left(2 \pi f_{c}t+\phi(t)\right)\right)\cdot \nonumber \\
   & \:\:\:\:\:\:\:\:\: \mathrm{min}\left(|A(t)\cos\left(2 \pi f_{c}t+\phi(t)\right)|,\! V_{sat}\right),
\end{align}
where $A(t)$, $f_c$ and $\phi(t)$ represent the amplitude, the carrier frequency and the phase of the PA input signal $x_p(t)$. If the input signal exceeds $V_{sat}$, the output signal $v_t(t)$ is set to $V_{sat}$. The signal $v_t(t)$ goes then through the BPF built by two capacitances and two inductances. This BPF has the impulse response $h_{\text{bpf}}(t)$ and the bandwidth $B_{\text{bpf}}$, which can be adjusted by the choice of the reactances. The PA output signal $y_p(t)$ is the band-pass filtered version of $v_t(t)$, $y_p(t) = h_{\text{bpf}}(t) \ast v_t(t)$, and can be given in the following way:
 \begin{align}
  y_{p}(t) &= R_{L}\cdot i_{L}(t) \approx f(A(t))\cos(2 \pi f_{c}t+\phi(t)), \nonumber \\
 \text{with } i_{L}(t) &\approx I_{L}(t)\cos(2 \pi f_{c}t+\phi(t)). 
  \label{eq:in_out_pa}
 \end{align}    
The resulting AM-AM transfer characteristic $f(\bullet)$ is a soft limiter type non-linearity and plotted in Fig. \ref{fig:PA_AM_AM}.

\subsubsection{Power consumption}
The power consumed from the battery $V_{sat}$ by the push-pull model can be calculated as \cite{mezghani2011modeling}:
   \begin{equation}\label{5}
   P_{\mathrm{PA}} = V_{sat}\cdot E[|i_{L}(t)|].
 \end{equation}
The output current  $i_{L}(t)$ is directly controlling $ P_{\mathrm{PA}}$. The output current $i_{L}(t)$ is a function of the input signal $x_p(t)$ as it is behaving as a voltage controlled current source before running into saturation region at high input values. Before concluding the PA section, one important parameter in which we are interested is the input back-off (IBO) of the push-pull model.
The IBO in push-pull model can be considered as the amount of clipping introduced in the signal \cite{mezghani2011modeling}. It can be given as:
\begin{equation}\label{4}
\text{IBO} =\frac{V_{sat}}{\sigma_{x_p}},
\end{equation}
where $\sigma_{x_p}^2$ is the variance of the input signal $x_p(t)$. ${\sigma_{x_p}}$ is measured in terms of voltage. The PA runs in the non-linear region if the IBO value is small as the saturation value $V_{sat}$ gets smaller than ${\sigma_{x_p}}$.

The average output power of the push-pull model $P_{\mathrm{T}}$, that is the power consumed by the antenna load $R_L$, is expressed as:
   \begin{equation}\label{3}
    P_{\mathrm{T}}= \text{E}[i_{L}(t)\cdot y_p(t)].
 \end{equation}

\medskip{}

\subsection{Power consumption of DAC/ADC}
The power consumptions of the ADC and DAC scale exponentially with the resolution as shown in \cite{Lee2008} and \cite{Cui2005}. The extreme case of $1$-bit ADC and DAC architecture consumes the least amount of power. Motivated by this fact, the optimal system parameters using $1$-bit ADC and DAC are investigated for reduced overall power consumption.

\section{Performance measures}
\label{sec:performance_measures}
Before formulating the problem, some performance measures have to be first introduced.
\subsection{Information rate $R$}
The information rate $R$ is given in bits per second and is defined as:
\begin{equation}
R = B \cdot I(X;\hat X),
\end{equation}
where $I(X;\hat X)$ is the mutual information between the sent and received signals $x[n]$ and $\hat x[n]$ defined as:
\begin{equation}
I(X;\hat X) = \sum_x \sum_{\hat x} P(x,\hat x) \log_2 \frac{P(x|\hat x)}{P(x)}, 
\end{equation}
where $P(\bullet)$ denotes the probability mass function (pmf). Memory effects of the filters are neglected. In the simulations, the mutual information is estimated based on the numerical method introduced in \cite{MI}, in order to take the different effects of communication chain into account, especially the nonlinear effects of the PA and the DAC/ADC.
\subsection{Signal bandwidth $B_{\mathrm{PA}}$}
The signal bandwidth $B_{\mathrm{PA}}$ designates the bandwidth of the transmitted signal at the output of the PA $y_p(t)$. Since the 3dB bandwidth definition does not inform us about the amount of OBR, another definition of the bandwidth is required. To this end, the signal-to-interference-noise ratio (SINR) is taken into account, wich is defined as SINR$=\alpha \frac{P_{\text{T}}}{\sigma^2_n+\sigma^2_i}$, where $\sigma^2_i$ denotes the interference power of both adjacent channels spaced by $B_{\mathrm{PA}}$ with the assumption that $\sigma^2_i = 2 \sigma^2_n$. For an SINR value of 10dB we get the 93,75\% bandwidth $B_{\mathrm{PA}}$:
\begin{align}
\frac{\int_{f_c -B_{\mathrm{PA}}/2}^{f_c+B_{\mathrm{PA}}/2} S(f) df}{\int_{- \infty}^{+ \infty} S(f) df} &= 93.75 \%,
\end{align}
where $S(f)$ is the power sprectral density of the PA output signal $y_p(t)$. The remaining $6.25\%$§ of the signal power lying outside the defined bandwidth $B_{\mathrm{PA}}$ (OBR) is then considered as $\sigma^2_i$.
\subsection{Transmit power efficiency $\eta_{\text{P}}$}
The transmit power efficiency $\eta_\text{P}$ measures the required PA power consumption for achieving the maximal information rate:
\begin{equation}
\eta_{\text{P}} = \frac{R}{P_{\mathrm{PA}}}.
\end{equation}
The higher $\eta_{\text{P}}$, the better the PA power is utilized.

\subsection{Transmit spectral efficiency $\eta_{\text{B}}$}
The transmit spectral efficiency $\eta_{\text{B}}$ measures the rate, at which information in bits per second can be transmitted in each Hz of bandwidth $B_{\mathrm{PA}}$:
\begin{equation}
\eta_{\text{B}} = \frac{R}{B_{\mathrm{PA}}}.
\end{equation}
The larger $\eta_{\text{B}}$, the better the bandwidth is utilized.

\section{Problem Formulation}
\label{sec:problem_formulation}
In this contribution, the aim is to minimize the total power consumption $P_{\text{Total}}$ of the wireless communication system through reducing the power consumed by the analog components $P_{\mathrm{Analog}}$. To this end, the power consumptions of the DAC, ADC and PA have to be reduced. The minimal power consumption of DAC and ADC can be reached by using 1-bit resolution. Thus, it remains to reduce the PA power consumption $P_{\mathrm{PA}}$.
Therefore, the goal of this work consists in finding the optimal PA parameters: IBO$_{\text{opt}}$ and $B_{\text{bpf,opt}}$, that maximize the transmit spectral effciency $\eta_{\text{B}}$ as well as the transmit power efficiency $\eta_{\text{P}}$ of a SISO wireless communication system with QPSK modulation and using 1-bit ADC/DAC. On one hand, increasing $\eta_{\text{P}}$ means reducing $P_{\mathrm{PA}}$ which implies reducing IBO. However, low IBO, i.e. low clipping ratio, results in a large amount of OBR. The latter increases the signal bandwidth $B_{\text{PA}}$ and thus decreases $\eta_{\text{B}}$. On the other hand, low $B_{\text{bpf}}$ can reduce the OBR which decreases $B_{\text{PA}}$ and thus increases $\eta_{\text{B}}$. Therefore, maximizing $\eta_{\text{B}}$ as well as maximizing $\eta_{\text{P}}$ are conflicting goals \cite{Nossek2008}. An optimal compromise is achieved by maximizing the product of both:
\begin{align}
\lbrace\text{IBO}_{\text{opt}},B_{\text{bpf,opt}}\rbrace &= \max_{\text{IBO}, B_{\text{bpf}}} \left\lbrace  \eta_{\text{P}} \cdot \eta_{\text{B}}\right\rbrace. 
\end{align}
The FOM is defined as: $\text{FOM} = \eta_{\text{P}} \cdot \eta_{\text{B}}$.
Since the problem only involves two variables, it can be solved based on numerical simulations using the Grid-Search method.
\section{Simulation Results}
\label{sec:simulation_results}
For the simulations, we choose a carrier frequency of $f_c=30 \cdot B$. The RRC filters have roll-off factor of $\rho=0.5$. The LPF is a 4th order Butterworth filter of 3-dB bandwidth equal to $B$. The PA includes a 4th order band-pass filter of 3-dB bandwidth $B_{\mathrm{bpf}}$ around $f_c$. The simulations are run with $N=10^4$ of QPSK symbols. The SINR is set to $10$dB. First simulations are done with sys2 and then all three systems are compared.

%
%
%
%

The aim is to find IBO$_{\text{opt}}$ and $B_{\text{bpf,opt}}$, that maximize the FOM, i.e. that jointly maximize the information rate $R$ and minimize the PA power consumption $P_{\mathrm{PA}}$ and the signal bandwidth $B_{\mathrm{PA}}$. Therefore, $R$, $P_{\mathrm{PA}}$ and $B_{\mathrm{PA}}$ are studied for different values of IBO and $B_{\text{bpf}}$. 

\begin{figure}
\centering
\psfrag{datadatadatadata1}[][]{$B_{\text{bpf}}=0.4 B$}
\psfrag{datadatadatadata2}[][]{$B_{\text{bpf}}=1 B$}
\psfrag{datadatadatadata3}[][]{$B_{\text{bpf}}=2 B$}
\psfrag{datadatadatadata4}[][]{$B_{\text{bpf}}=10 B$}
\psfrag{power}[][]{$P_{\mathrm{PA}}/P_{\mathrm{T}}$}
\psfrag{IBO}[][]{IBO}      
\epsfig{file= 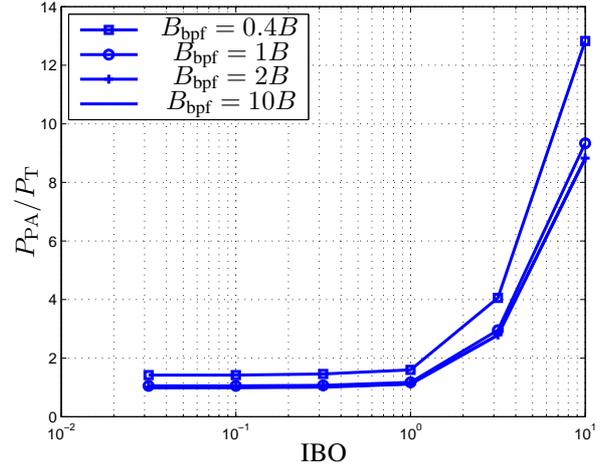, width = \columnwidth}
\caption{Inverse PA efficiency $P_{\mathrm{PA}}/P_{\mathrm{T}}$ vs. IBO for different $B_{\text{bpf}}$}
\label{P_ibo_bpf}
\end{figure}

\begin{figure}    
\centering
\psfrag{datadatadatadata1}[][]{$B_{\text{bpf}}=0.4 B$}
\psfrag{datadatadatadata2}[][]{$B_{\text{bpf}}=1 B$}
\psfrag{datadatadatadata3}[][]{$B_{\text{bpf}}=2 B$}
\psfrag{datadatadatadata4}[][]{$B_{\text{bpf}}=10 B$}
\psfrag{R/B}[][]{$R/B$}      
\psfrag{IBO}[][]{IBO}      
\epsfig{file= 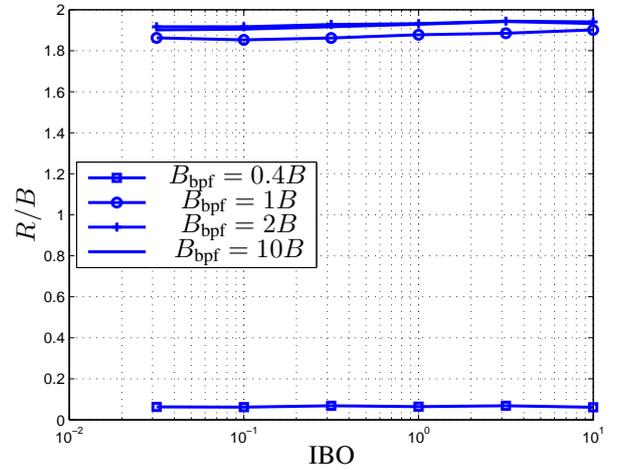, width = \columnwidth}
\caption{Information rate $R/B$ in bits per channel use vs. IBO for different $B_{\text{bpf}}$}
\label{R_ibo_bpf}
\end{figure}

\begin{figure}
\centering
\psfrag{datadatadatadata1}[][]{$B_{\text{bpf}}=0.4 B$}
\psfrag{datadatadatadata2}[][]{$B_{\text{bpf}}=1 B$}
\psfrag{datadatadatadata3}[][]{$B_{\text{bpf}}=2 B$}
\psfrag{datadatadatadata4}[][]{$B_{\text{bpf}}=10 B$}
\psfrag{bandwidth}[][]{$B_{\mathrm{PA}}/B$}      
\psfrag{IBO}[][]{IBO}      
\epsfig{file= 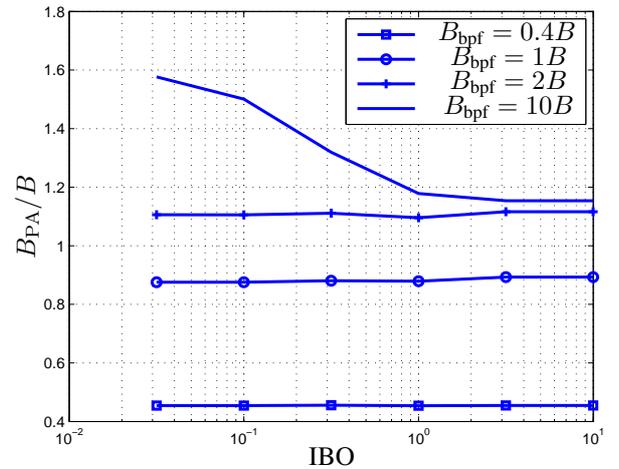, width = \columnwidth}
\caption{Normalized signal bandwidth $B_{\mathrm{PA}}/B$ vs. IBO for different $B_{\text{bpf}}$}
\label{B_ibo_bpf}
\end{figure}

In Fig. \ref{P_ibo_bpf}, the inverse of the PA efficiency $P_{\mathrm{PA}}/P_{\text{T}}$ is plotted as function of IBO. This increases exponentially with IBO. For low IBO, the clipping ratio is very small. The PA operates in the non-linear region. Therefore, the PA power consumption $P_{\text{PA}}$ decreases. So, it is interesting to run the PA at low IBO to reduce $P_{\mathrm{PA}}$. However, the effect of low IBO at $R$ and $B_{\mathrm{PA}}$ for different $B_{\text{bpf}}$ values has to be investigated.
In the simulation set depicted in Fig. \ref{R_ibo_bpf} and Fig. \ref{B_ibo_bpf}, the information rate and the signal bandwidth normalized by $B$, $R/B$ and $B_{\mathrm{PA}}/B$ are plotted as functions of IBO for different PA BPF bandwidths $B_{\text{bpf}}$. The lower IBO, the more the PA works in the non-linear region. That means the PA input signal gets more clipped which increases the OBR at the PA output. The amount of OBR can be restricted by the choice of $B_{\text{bpf}}$. As can be seen in the simulation results, for low IBO the $R/B$ decreases slightly while $B_{\mathrm{PA}}/B$ increases. However, the decrease in $R/B$ and the increase in $B_{\mathrm{PA}}/B$ depends on the used BPF. The larger $B_{\mathrm{bpf}}$, the less the removed OBR and thus the larger $B_{\mathrm{PA}}/B$. The decrease in $B_{\mathrm{bpf}}$ leads to a decrease in $B_{\mathrm{PA}}/B$. However, the further decrease in $B_{\mathrm{bpf}}$ leads to a decrease in $R/B$. Therefore, $B_{\mathrm{bpf}}$ has to be optimized to increase the FOM. 

\begin{figure}
\centering
\psfrag{datadatadatadata1}[][]{$B_{\text{bpf}}=0.4 B$}
\psfrag{datadatadatadata2}[][]{$B_{\text{bpf}}=1 B$}
\psfrag{datadatadatadata3}[][]{$B_{\text{bpf}}=2 B$}
\psfrag{datadatadatadata4}[][]{$B_{\text{bpf}}=10 B$}
\psfrag{FOM}[][]{FOM $\cdot N_0/\alpha$ }      
\psfrag{IBO}[][]{IBO}      
\epsfig{file= 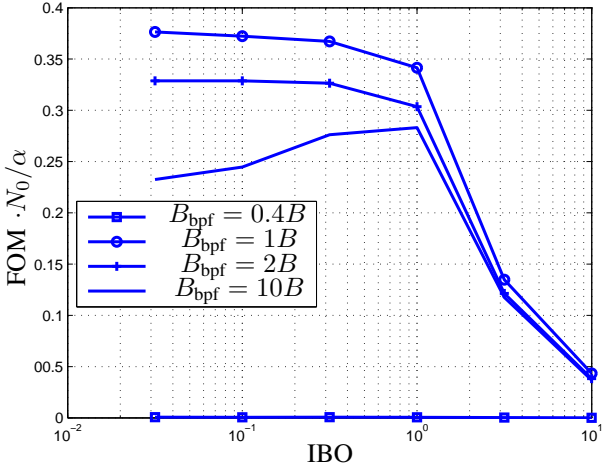, width = \columnwidth}
\caption{Normalized FOM ($\eta_{\text{P}} \cdot \eta_{\text{B}}$) vs. IBO for different $B_{\text{bpf}}$}
\label{fom_ibo_bpf}
\end{figure}

\begin{figure}
\centering
\psfrag{datadatadatadatadata1}[][]{IBO $= 0.0316$}
\psfrag{datadatadatadatadata2}[][]{IBO $= 0.1$}
\psfrag{datadatadatadatadata3}[][]{IBO = $1$}
\psfrag{datadatadatadatadata4}[][]{IBO = $10$}
\psfrag{FOM}[][]{FOM $\cdot N_0/\alpha$ }      
\psfrag{bpf}[][]{$B_{\text{bpf}}/B$}      
\epsfig{file= 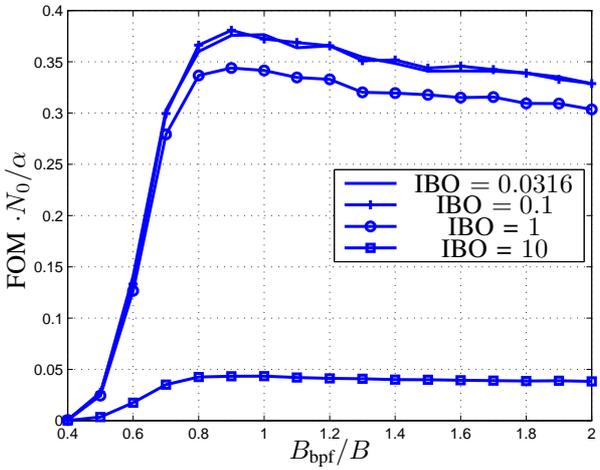, width = \columnwidth}
\caption{Normalized FOM ($\eta_{\text{P}} \cdot \eta_{\text{B}}$) vs. $B_{\text{bpf}}/B$ for different IBO values}
\label{fom_bpf}
\end{figure}

In Fig.\ref{fom_ibo_bpf}, the FOM is plotted as function of IBO for different $B_{\mathrm{bpf}}$. The FOM is maximized and saturated for low IBO and starts decreasing from IBO=1. The maximal value of FOM depends on $B_{\mathrm{bpf}}$. It can be concluded that there is an optimal $B_{\mathrm{bpf}}$ that maximizes the FOM at low IBO. The optimal value of $B_{\mathrm{bpf}}$ can be extracted from Fig. \ref{fom_bpf}. Thus, IBO$_{\text{opt}} \leq 0.1$ and $B_{\text{bpf,opt}}=0.9 B$ are the optimal operating points that solve the optimization problem for sys2.

\begin{figure}
\centering
\psfrag{datadata1}[][]{sys1}
\psfrag{datadata2}[][]{sys2}
\psfrag{datadata3}[][]{sys3}
\psfrag{FOM}[][]{FOM $\cdot N_0/\alpha$ }      
\psfrag{bpf}[][]{$B_{\text{bpf}}/B$}      
\epsfig{file= 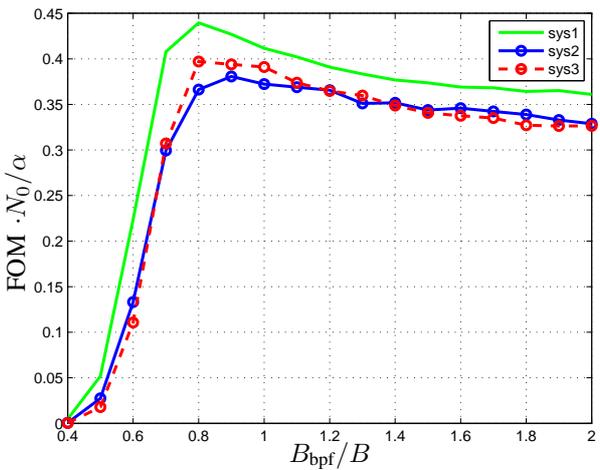, width = \columnwidth}
\caption{Normalized FOM ($\eta_{\text{P}} \cdot \eta_{\text{B}}$) vs. $B_{\text{bpf}}/B$ for IBO$=0.1$}
\label{fom_eta_3syss}
\end{figure}

In the last simulation set, all three systems sys1, sys2 and sys3 introduced in section \ref{sec:system_model} are considered. In Fig. \ref{fom_eta_3syss}, the FOM is plotted as function of $B_{\text{bpf}}$ for the three introduced system models at IBO$ = 0.1$. Sys1 is used as the benchmark. As can be seen from the figure, the FOM is maximized at $B_{\text{bpf}} = 0.8 B, 0.9 B,0.8 B$ for sys1, sys2 and sys3, respectively. The simulation results show that there is a loss in FOM of sys2 compared to sys1. This loss is due to the reduction of the DAC/ADC resolution to 1-bit. However, almost half of this loss can be recovered in sys3 by removing the RRC filter before the 1-bit DAC at the transmitter. In sys2, the QPSK symbols are pulse shaped by an RRC filter and then quantized by the 1-bit DAC. The output of the 1-bit DAC is again QPSK. So what is the need for the pulse shaper? The idea with sys3 is to keep the QPSK symbols without digital signal processing so that the output of the 1-bit DAC consists of the same QPSK symbols but in the analog domain. There is no quantization loss in this case. 
Note that the total power consumption $P_{\text{Total}}$ of sys3 is largely decreased compared to $P_{\text{Total}}$ of sys1, since 1-bit DAC/ADC are used and the RRC filter at the transmitter is removed.


\section{Conclusion}
\label{sec:conclusion}
In this work, it is shown that the PA can be run in the non-linear region with an appropriate BPF while maximizing the power and spectral efficency in the SISO system with the QPSK modulation scheme and using 1-bit quantization. Additionally, the removal of the RRC filter at the transmitter side is beneficial in terms of the introduced FOM. The FOM gets improved and the circuit complexity at the transmitter is decreased.
For future work, one can look for the optimal analog pulse shaper at the transmitter side in the presence of 1-bit DAC. Furthermore, the same performance analysis can be investigated while considering the AM-PM distortion of the PA. Moreover, the loss due to the 1-bit quantization can be recovered by oversampling in time and in space. Then, the analysis can be generalized to the MIMO case.


\bibliographystyle{IEEEtran}

\bibliography{Paper}

\end{document}